\documentclass[12pt,thmsa]{article}
\usepackage{sw20aip}

\input tcilatex
\begin{document}

\title{Path integral treatment of a family of super-integrable systems in $n$%
-dimensional Euclidean space}
\author{M. T. Chefrour \\
D\'epartement de Physique, Facult\'e des Sciences,\\
Universit\'e Badji Mokhtar, Annaba, Alg\'eria. \and F. Benamira and L.
Guechi \\
Laboratoire de Physique Th\'eorique,\\
D\'epartement de Physique, Facult\'e des Sciences,\\
Universit\'e Mentouri, Route d'Ain El Bey, \\
Constantine, Alg\'eria. \and S. Mameri \\
D\'epartement de Physique, Facult\'e des Sciences,\\
Universit\'e Ferhat Abbas, Setif, Alg\'eria.}
\maketitle

\begin{abstract}
The exact path integration for a family of maximally super-integrable
systems generalizing the hydrogen atom in the $n$-dimensional Euclidean
space is presented. The Green's function is calculated in parabolic
rotational and spherical coordinate systems. The energy spectrum and the
correctly normalized wave functions of the bound states are obtained from
the poles of the Green's function and their residues, respectively.



\end{abstract}

\section{Introduction}

During the last 40 years, super-integrable dynamical systems have been the
object of a considerable interest for their pratical importance in various
fields. A general classification of Hamiltonians in two and three dimensions
possessing dynamical invariance groups was initiated by Smorodinsky,
Winternitz and collaborators\cite{JASPW} , continued by Kibler and Winternitz%
\cite{MKPW} , and revived, in recent years, by Evans\cite{NWE1} . This
classification was established according to the number of degrees of
freedom, quadratic integrals of motion in the momenta and coordinate systems
in which the potential allows the separation of variables. The Hamiltonian
systems with these potentials are called super-integrable. Generally, in $n$
dimensions, a system is called ''minimally'' super-integrable if it has $%
(2n-2)$ constants or integrals of motion ( including energy ), and it is
called ''maximally'' super-integrable if it has $(2n-1)$ integrals of motion%
\cite{JH,DBCDKK} . A list of potentials corresponding to minimally
super-integrable and maximally super-integrable systems with corresponding
constants of motion in the classical form and all separating coordinate
systems has been established by Evans\cite{NWE2} .

In the last decade, Grosche et al have presented a detailed path integral
discussion of the so-called Smorodinsky-Winternitz super-integrable
Hamiltonians in the three further spaces of constant curvature: two- and
three-dimensional Euclidean space \cite{CGGSPANS1} , two- and
three-dimensional sphere\cite{CGGSPANS2} , two- and three-dimensional
hyperboloid \cite{CGGSPANS3,CGGSPANS4} . Chetouani et al \cite{LCLGTFH1} ,
whose work followed closely upon that of Grosche and co-workers studied
three classes of two- and three-dimensional Smorodinsky-Winternitz
super-integrable Hamiltonian systems in Euclidean space using a variant of
the so(2,1) algebraic approach.

Recently, a new family of maximally super-integrable systems generalizing
the hydrogen atom has been constructed in $n$-dimensional Euclidean space by
Winternitz and Rodriguez \cite{PWMAR} and studied in standard quantum
mechanics through resolution of Schr\"odinger equation in parabolic
rotational and spherical coordinates. This new family corresponds to the
potentials 
\begin{equation}
V(x^{(1)},x^{(2)},...,x^{(n)})=-\frac \gamma r+\stackunder{i=1}{\stackrel{n-1%
}{\sum }}\frac{\beta _i}{\left( x^{(i)}\right) ^2},  \label{a.1}
\end{equation}
where $\gamma $ and $\beta _i$ are positive constants and $r=\left(
x^{(1)2}+x^{(2)2}+...x^{(n)2}\right) ^{\frac 12}.$

In this paper, we want to present a path integral treatment of these
potentials in parabolic rotational and spherical coordinate systems.

our work can be seen as a generalization of the earlier work of Chetouani
and Hammann \cite{LCTFH} , in which they established only the $n$%
-dimensional Coulomb Green function in polar coordinates by means of the
Gegenbauer expansion formula \cite{ISGIMR}.

Our study is organized as follows: in the following section, we introduce $n$%
-dimensional parabolic rotational coordinates and formulate the path
integral associated with the potential $V(x^{(1)},x^{(2)},...,x^{(n)})$. An
explicit Lagrangian path integral formulation is derived. By means of
successive applications of appropriate time transformations we reduce the $%
\left\{ \theta ^{(k)}\right\} $ and $(\xi ,\eta )$ path integrations to the
already known ones. The Green function is evaluated in a spectral
decomposition. In section III, we construct the Green function in $n$%
-dimensional spherical coordinates. The energy spectrum and the normalized
wave functions are found as in the case of parabolic rotational coordinates.
The section IV will be a conclusion.

\section{Parabolic rotational coordinates in E$_n$ and path integral}

In $n$-dimensional parabolic rotational coordinates

\begin{eqnarray}
x^{(1)} &=&\xi \eta \cos \theta ^{(1)}\cos \theta ^{(2)}...\cos \theta
^{(n-3)}\cos \theta ^{(n-2)},  \nonumber  \label{a.1} \\
x^{(2)} &=&\xi \eta \cos \theta ^{(1)}\cos \theta ^{(2)}...\cos \theta
^{(n-3)}\sin \theta ^{(n-2)},  \nonumber  \label{a.2} \\
x^{(3)} &=&\xi \eta \cos \theta ^{(1)}\cos \theta ^{(2)}...\sin \theta
^{(n-3)},  \nonumber \\
&&\ .  \nonumber \\
&&\ .  \nonumber \\
&&\ .  \nonumber \\
x^{(n-2)} &=&\xi \eta \cos \theta ^{(1)}\sin \theta ^{(2)},  \nonumber \\
x^{(n-1)} &=&\xi \eta \sin \theta ^{(1)},  \nonumber \\
x^{(n)} &=&\frac 12(\xi ^2-\eta ^2),  \label{a.2}
\end{eqnarray}
where $0\leq \xi ,\eta <\infty ,$ $0\leq \theta ^{(\nu )}<\pi $ ($\nu
=2,...,n-2),$ $0\leq \theta ^{(1)}<2\pi ,$ the classical Lagrangian reads as

\begin{eqnarray}
\mathcal{L}_{cl} &=&\,\,\frac M2\left[ \left( \xi ^2+\eta ^2\right) \left( 
\stackrel{.}{\xi }^2+\stackrel{.}{\eta }^2\right) +\xi ^2\eta ^2\left( (%
\stackrel{.}{\theta }^{(1)})^2+\cos ^2\theta ^{(1)}(\stackrel{.}{\theta }%
^{(2)})^2+...\right. \right.  \nonumber  \label{a.2} \\
&&\ \left. \left. +\cos ^2\theta ^{(1)}...\cos ^2\theta ^{(n-3)}(\stackrel{.%
}{\theta }^{(n-2)})^2\right) \right] -V,  \label{a.3}
\end{eqnarray}
where the potential $V(x^{(1)},x^{(2)},...,x^{(n)})$ can be written in the
form

\begin{eqnarray}
V &=&-\frac{2\gamma }{\xi ^2+\eta ^2}+\frac{\hbar ^2}{2M\xi ^2\eta ^2}\left[ 
\frac{p_1(p_1-1)}{\cos ^2\theta ^{(1)}...\cos ^2\theta ^{(n-2)}}+\frac{%
p_2(p_2-1)}{\cos ^2\theta ^{(1)}...\sin ^2\theta ^{(n-2)}}+...\right. 
\nonumber  \label{a.3} \\
&&\ \left. +\frac{p_{n-2}(p_{n-2}-1)}{\cos ^2\theta ^{(1)}\sin ^2\theta
^{(2)}}+\frac{p_{n-1}(p_{n-1}-1)}{\sin ^2\theta ^{(1)}}\right] ,  \label{a.4}
\end{eqnarray}
with 
\begin{equation}
p_i=\frac 12\pm \sqrt{\frac 14+\frac{2M}{\hbar ^2}\beta _i},\text{ }%
i=1,...,n-1.  \label{a.5}
\end{equation}

Hence the metric tensor in parabolic rotational coordinates has the form 
\begin{equation}
(g_{ab})=diag(g_{\xi \xi },g_{\eta \eta },g_{\theta ^{(1)}\theta
^{(1)}},...,g_{\theta ^{(n-2)}\theta ^{(n-2)}}),  \label{a.6}
\end{equation}
where

\begin{eqnarray}
g_{\xi \xi } &=&g_{\eta \eta }=\xi ^2+\eta ^2,\qquad g_{\theta ^{(1)}\theta
^{(1)}}=\xi ^2\eta ^2  \nonumber  \label{a.3} \\
g_{\theta ^{(i)}\theta ^{(i)}} &=&\xi ^2\eta ^2\stackunder{k=1}{\stackrel{i-1%
}{\prod }}\cos ^2\theta ^{(k)};\text{ }i=2,...,n-2.  \label{a.7}
\end{eqnarray}
The infinitesimal volume element $d^nx$ transforms as follows:

\begin{equation}
d^nx=\sqrt{g}d\xi d\eta d\theta ^{(1)}...d\theta ^{(n-2)},  \label{a.8}
\end{equation}
where 
\begin{equation}
\sqrt{g}=\sqrt{\det (g_{ab})}=(\xi ^2+\eta ^2)(\xi \eta )^{n-2}\stackrel{n-2%
}{\stackunder{k=1}{\prod }}\left( \cos \theta ^{(k)}\right) ^{n-2-k}.
\label{a.9}
\end{equation}
The inverse of (\ref{a.6}) is given by 
\begin{equation}
(g^{ab})=(g_{ab})^{-1}=diag(g_{\xi \xi }^{-1},g_{\eta \eta }^{-1},g_{\theta
^{(1)}\theta ^{(1)}}^{-1},...,g_{\theta ^{(n-2)}\theta ^{(n-2)}}^{-1}).
\label{a.10}
\end{equation}
The momentum operators are 
\begin{eqnarray}
P_\xi &=&\frac \hbar i\left( \frac \partial {\partial \xi }+\frac{\Gamma
_\xi }2\right) ;\qquad \Gamma _\xi =\frac{2\xi }{\xi ^2+\eta ^2}+\frac{n-2}%
\xi ,  \nonumber \\
P_\eta &=&\frac \hbar i\left( \frac \partial {\partial \eta }+\frac{\Gamma
_\eta }2\right) ;\qquad \Gamma _\eta =\frac{2\eta }{\xi ^2+\eta ^2}+\frac{n-2%
}\eta ,  \nonumber \\
P_{\theta ^{(k)}} &=&\frac \hbar i\left( \frac \partial {\partial \theta
^{(k)}}+\frac{\Gamma _{\theta ^{(k)}}}2\right) ;\Gamma _{\theta
^{(k)}}=-(n-2-k)\tan \theta ^{(k)},k=1,...,n-2.  \nonumber  \label{a.7} \\
&&  \label{a.11}
\end{eqnarray}

In order to express the quantum Hamiltonian by position and momentum
operators, we can choose the Weyl ordering prescription \cite{MM,TDL,CGFS} ,
or the product-ordering prescription \cite{CG} . Due to the special nature
of $(g_{ab})$ and $(g^{ab})$ respectively, the quantum correction $\triangle
V$ is the same for both prescriptions. Consequently, we obtain for the
Hamiltonian

\begin{eqnarray}
H &=&\frac 1{2M}\left[ \frac 1{\sqrt{\xi ^2+\eta ^2}}\left( P_\xi ^2+P_\eta
^2\right) \frac 1{\sqrt{\xi ^2+\eta ^2}}+\frac 1{\xi ^2\eta ^2}\left[
P_{\theta ^{(1)}}^2+\right. \right.   \nonumber \\
&&\left. \left. \left. \frac 1{\cos ^2\theta ^{(1)}}\left[ P_{\theta
^{(2)}}^2+\frac 1{\cos ^2\theta ^{(2)}}\left[ P_{\theta ^{(3)}}^2+...+\frac
1{\cos ^2\theta ^{(n-3)}}P_{\theta ^{(n-2)}}^2\right] \right. ...\right]
\right] \right]   \nonumber \\
&&-\frac{2\gamma }{\xi ^2+\eta ^2}+\frac{\hbar ^2}{2M\xi ^2\eta ^2}\left[ 
\frac{p_1(p_1-1)}{\cos ^2\theta ^{(1)}...\cos ^2\theta ^{(n-2)}}+\frac{%
p_2(p_2-1)}{\cos ^2\theta ^{(1)}...\sin ^2\theta ^{(n-2)}}+\right.  
\nonumber \\
&&\left. ...+\frac{p_{n-2}(p_{n-2}-1)}{\cos ^2\theta ^{(1)}\sin ^2\theta
^{(2)}}+\frac{p_{n-1}(p_{n-1}-1)}{\sin ^2\theta ^{(1)}}\right] +\triangle V,
\label{a.12}
\end{eqnarray}
with the quantum potential $\triangle V$ given by

\begin{eqnarray}
\triangle V &=&\frac{\hbar ^2}{8M\xi ^2\eta ^2}\left[ (n-2)(n-4)+(n-3)^2\tan
^2\theta ^{(1)}-\frac{2(n-3)}{\cos ^2\theta ^{(1)}}\right.  \nonumber \\
&&+\frac 1{\cos ^2\theta ^{(1)}}\left[ (n-4)^2\tan ^2\theta ^{(2)}-\frac{%
2(n-4)}{\cos ^2\theta ^{(2)}}\right.  \nonumber \\
&&+\frac 1{\cos ^2\theta ^{(2)}}\left[ (n-5)^2\tan ^2\theta ^{(3)}-\frac{%
2(n-5)}{\cos ^2\theta ^{(3)}}+...+\right.  \nonumber \\
&&\left. \left. \left. \frac 1{\cos ^2\theta ^{(n-4)}}\left[ \tan ^2\theta
^{(n-3)}-\frac 2{\cos ^2\theta ^{(n-3)}}\right] ...\right] \right] \right] .
\label{a.13}
\end{eqnarray}

In manner similar to that in Refs.\cite{CGFS,CG} , by direct calculation we
get for the Lagrangian path integral in the ''product form''-definition

\begin{eqnarray}
K(\stackrel{\rightarrow }{r}_f,\stackrel{\rightarrow }{r}_i;T)\!\!\!
&=&\!\!\!\!\!\!\int \!\!\sqrt{g}\mathcal{D}\xi (t)\mathcal{D}\eta (t)%
\mathcal{D}\theta ^{(1)}(t)...\mathcal{D}\theta ^{(n-2)}(t)\exp \left\{
\frac i\hbar \!\int_0^T\!\left( \mathcal{L}_{cl}\!-\!\triangle V\right)
dt\right\}  \nonumber \\
\!\!\!\!\!\!\!\! &=&\!\stackunder{N\rightarrow \infty }{\lim }\int \!\!%
\stackunder{j=1}{\stackrel{N}{\prod }}\!\!\!\left( \frac M{2i\pi \hbar
\varepsilon }\right) ^{\frac n2}\!\stackunder{j=1}{\stackrel{N-1}{\prod }%
\!\!\!}\left( \xi _j^2+\eta _j^2\right) \left( \xi _j\eta _j\right) ^{n-2}%
\stackrel{n-2}{\stackunder{k=1}{\prod }}\!\!\left( \cos \theta
_j^{(k)}\right) ^{n-2-k}  \nonumber \\
&&\times d\xi _jd\eta _jd\theta _j^{(1)}...d\theta _j^{(n-2)}\exp \left\{
\frac i\hbar \stackrel{N}{\stackunder{j=1}{\sum }}S^{pc}(j,j-1)\right\} ,
\label{a.14}
\end{eqnarray}
with the short-time action 
\begin{eqnarray}
S^{pc}(j,j-1) &=&\frac M{2\varepsilon }\left[ \widehat{\left( \xi _j^2+\eta
_j^2\right) }\left( \triangle ^2\xi _j+\triangle ^2\eta _j\right) +\widehat{%
\xi _j^2\eta _j^2}\left( \triangle ^2\theta _j^{(1)}\right. \right. 
\nonumber \\
&&\left. \left. +\widehat{\cos ^2\theta _j^{(1)}}\triangle ^2\theta
_j^{(2)}+...+.\widehat{\cos ^2\theta _j^{(1)}}...\widehat{\cos ^2\theta
_j^{(n-3)}}\triangle ^2\theta _j^{(n-2)}\right) \right]  \nonumber \\
&&+\frac{2\varepsilon \gamma }{\xi _j^2+\eta _j^2}-\frac{\varepsilon \hbar ^2%
}{2M\xi _j^2\eta _j^2}\left[ -\frac{(n-2)(n-4)}4-\frac{(n-3)^2}4\right. 
\nonumber \\
&&+\frac{p_{n-1}(p_{n-1}-1)}{\sin ^2\theta _j^{(1)}}+\frac{(n-3)(n-5)}{4\cos
^2\theta _j^{(1)}}  \nonumber \\
&&+\frac 1{\cos ^2\theta _j^{(1)}}\left[ -\frac{(n-4)^2}4+\frac{(n-4)(n-6)}{%
4\cos ^2\theta _j^{(2)}}+\frac{p_{n-2}(p_{n-2}-1)}{\sin ^2\theta _j^{(2)}}%
\right.  \nonumber \\
&&+\!\frac 1{\cos ^2\theta _j^{(2)}}\left[ \!-\frac{(n-5)^2}4\!+\!\frac{%
(n-5)(n-7)}{4\cos ^2\theta _j^{(3)}}\!+\!\frac{p_{n-3}(p_{n-3}\!-\!1)}{\sin
^2\theta _j^{(3)}}\!+\!...\right.  \nonumber \\
&&\left. \left. \left. +\frac 1{\cos ^2\theta _j^{(n-3)}}\left[ \frac{%
p_2(p_2-1)}{\sin ^2\theta _j^{(n-2)}}+\frac{p_1(p_1-1)}{\cos ^2\theta
_j^{(n-2)}}\right] ...\right] \right] \right] .  \label{a.15}
\end{eqnarray}
In the above we have used the notations: $\varepsilon =t_j-t_{j-1},$ $%
T=N\varepsilon =t_f-t_i,$ $\stackrel{\rightarrow }{r}_f=\stackrel{%
\rightarrow }{r}(t_N),$ $\stackrel{\rightarrow }{r}_i=\stackrel{\rightarrow 
}{r}(t_0),$ $\widehat{\xi _j^2+\eta _j^2}=\sqrt{\left( \xi _j^2+\eta
_j^2\right) \left( \xi _{j-1}^2+\eta _{j-1}^2\right) };\quad \widehat{\xi
_j^2\eta _j^2}=\xi _j\eta _j\xi _{j-1}\eta _{j-1;\quad }\widehat{\cos
^2\theta _j^{(k)}}=\cos \theta _j^{(k)}\cos \theta _{j-1}^{(k)}.$

First, the $\left\{ \theta \right\} $ path integrations can be entangled
with one another by means of pure and appropriate time transformations \cite
{IHDHK,AI} $t\rightarrow s$ defined by $dt=\xi ^2\eta ^2\stackrel{n-2}{%
\stackunder{k=i}{\prod }}\cos ^2\theta ^{(n-2-k)}ds;(i=1,2,...n-3),$ and
carried out using the P\"oschl-Teller path integral solution $(0<\theta
<\frac \pi 2)$ \cite{MBGJ,IHD,CGFS1,WFHLPM,AIHKCCG,HKIM} 
\begin{eqnarray}
K^{(PT)}(\theta _f,\theta _i;T)\!\! &=&\!\!\!\!\int \mathcal{D}\theta
(t)\exp \left\{ \frac i\hbar \int_0^T\left[ \frac M2\stackrel{.}{\theta }%
^2\!-\frac{\hbar ^2}{2M}\left( \frac{\kappa ^2-\frac 14}{\cos ^2\theta }+%
\frac{\lambda ^2-\frac 14}{\sin ^2\theta }\right) \!\right] dt\right\} 
\nonumber \\
\ &=&\!\stackunder{J_n=0}{\stackrel{\infty }{\sum }}\exp \left\{ -\frac{%
i\hbar T}{2M}(\kappa +\lambda +2J_n+1)^2\right\} \!\Psi _{J_n}^{(\lambda
,\kappa )*}(\theta _i)\Psi _{J_n}^{(\lambda ,\kappa )}(\theta _f),  \nonumber
\label{a.12} \\
&&  \label{a.16}
\end{eqnarray}
with the normalized wave functions given by

\begin{eqnarray}
\Psi _{J_n}^{(\lambda ,\kappa )}(\theta ) &=&\left[ 2(\kappa +\lambda
+2J_n+1)\frac{J_n!\Gamma (\kappa +\lambda +J_n+1)}{\Gamma (\kappa
+J_n+1)\Gamma (\lambda +J_n+1)}\right] ^{\frac 12}  \nonumber  \label{a.13}
\\
&&\times (\sin \theta )^{\lambda +\frac 12}(\cos \theta )^{\kappa +\frac
12}P_{J_n}^{(\lambda ,\kappa )}(\cos 2\theta ),  \label{a.17}
\end{eqnarray}
where $P_{J_n}^{(\lambda ,\kappa )}(\cos 2\theta )$ are Jacobi polynomials (%
\cite{ISGIMR} , p.1035). After performing the integrations over all the
angular variables $\theta _j^{(k)}$ $(k=1,...,n-2),$ we obtain for the
Feynman propagator (\ref{a.14}) the following expansion:

\begin{equation}
\!\!\!\!\!\!\!\!K(\stackrel{\rightarrow }{r}_f,\stackrel{\rightarrow }{r}%
_i;T)\!\!=\stackrel{\infty }{\stackunder{\left\{ J_{n-2}\right\} =0}{\sum }}%
\!\!\Psi _{\left\{ J_{n-2}\right\} }^{*}\!\left( \!\left\{ \theta
_i^{(n-2)}\right\} \!\right) \!\Psi _{\left\{ J_{n-2}\right\} }\!\left(
\!\left\{ \theta _f^{(n-2)}\right\} \!\right) \!\!K_{\left\{ J_{n-2}\right\}
}(\xi _f,\eta _f,\xi _i,\eta _i;T),  \label{a.18}
\end{equation}
where we have abbreviated for $\left\{ J_{n-2}\right\} $ and $\left\{ \theta
^{(n-2)}\right\} $ the sets $\left\{ J_1,J_2,...,J_{n-2}\right\} $ and $%
\left\{ \theta ^{(1)},\theta ^{(2)},...,\theta ^{(n-2)}\right\} ,$
respectively. The angular part of the wave functions is found to be

\begin{equation}
\Psi _{\left\{ J_n\right\} }\left( \left\{ \theta ^{(n-2)}\right\} \right) =%
\stackunder{l=1}{\stackrel{n-2}{\prod }}\Psi _{J_l}\left( \theta
^{(n-l-1)}\right) ,  \label{a.19}
\end{equation}
where 
\begin{eqnarray}
\Psi _{J_l}\left( \theta ^{(n-l-1)}\right)  &=&\left[ 2\frac{%
(m_{l-1}+p_{l+1}+2J_l+\frac 12)J_l!\Gamma (m_{l-1}+p_{l+1}+J_l+\frac 12)}{%
\Gamma (m_{l-1}+J_l+1)\Gamma (p_{l+1}+J_l+\frac 12)}\right] ^{\frac 12} 
\nonumber \\
&&\ \ \times (\sin \theta ^{(n-l-1)})^{p_{l+1}}(\cos \theta
^{(n-l-1)})^{m_{l-1}-\frac l2+1}  \nonumber  \label{a.19} \\
&&\ \ \times P_{J_l}^{(p_{l+1}-\frac 12,m_{l-1})}(\cos 2\theta ^{(n-l-1)}),
\label{a.20}
\end{eqnarray}
with 
\begin{equation}
m_0=p_1-\frac 12,\quad m_l=\stackunder{i=1}{\stackrel{l+1}{\sum }}p_i+2%
\stackunder{i=1}{\stackrel{l}{\sum }}J_i+\frac{l-1}2,\quad 1\leq l\leq n-2.
\label{a.21}
\end{equation}
The kernel $K_{\left\{ J_{n-2}\right\} }(T)$ is given by 
\begin{eqnarray}
K_{\left\{ J_{n-2}\right\} }(\xi _f,\eta _f,\xi _i,\eta _i;T)
&=&\!\!\!\left( \xi _f\xi _i\eta _f\eta _i\right) ^{\frac{2-n}2}\left(
4r_fr_i\right) ^{-\frac 12}\stackunder{N\rightarrow \infty }{\lim }\int 
\stackrel{N}{\stackunder{j=1}{\prod }}\!\!\left[ \!\frac{M\sqrt{4r_jr_{j-1}}%
}{2i\pi \hbar \varepsilon }\!\right]   \nonumber \\
&&\times \stackrel{N-1}{\stackunder{j=1}{\prod }}d\xi _jd\eta _j\exp \left\{
\frac i\hbar \stackrel{N}{\stackunder{j=1}{\sum }}A^{pc}(j,j-1)\right\} ,
\label{a.22}
\end{eqnarray}
with the short-time action 
\begin{equation}
A^{pc}(j,j-1)=\frac M{2\varepsilon }\left( \widetilde{\xi }_j^2+\widetilde{%
\eta }_j^2\right) \left( \triangle ^2\xi _j+\triangle ^2\eta _j\right) +%
\frac{2\varepsilon \gamma }{\widetilde{\xi }_j^2+\widetilde{\eta }_j^2}-%
\frac{\varepsilon \hbar ^2}{2M}\frac{(\sigma +\frac{n-3}2)^2-\frac 14}{%
\widetilde{\xi }_j^2\widetilde{\eta }_j^2},  \label{a.23}
\end{equation}
where we have set 
\begin{equation}
\sigma =\stackunder{i=1}{\stackrel{n-1}{\sum }}p_i+2\stackunder{i=1}{%
\stackrel{n-2}{\sum }}J_i,  \label{a.24}
\end{equation}
and $\widetilde{u}_j=(u_j+u_{j-1})/2$. This brings us to path integration
over the coordinates $\xi $ and $\eta .$ We obtain by introducing the energy 
$E$ with the help of the Green's function ( Fourier transform of the
propagator ), performing the time transformation $dt=(\xi ^2+\eta ^2)ds$ and
applying the path integral solution of the radial harmonic oscillator\cite
{DPAI,LCLGTFH} in the $\xi $ and $\eta $ variable, respectively: 
\begin{eqnarray}
\!\!\!\!\!\!\!\!\!\!\!\!\!G(\stackrel{\rightarrow }{r}_f,\stackrel{%
\rightarrow }{r}_i;E) &=&\int_0^\infty dT\exp (\frac i\hbar ET)K(\stackrel{%
\rightarrow }{r}_f,\stackrel{\rightarrow }{r}_i;T)  \nonumber \\
\!\!\!\!\!\!\!\!\!\!\!\!\! &=&\stackrel{\infty }{\stackunder{\left\{
J_{n-2}\right\} =0}{\sum }}\Psi _{\left\{ J_{n-2}\right\} }^{*}\left(
\left\{ \theta _i^{(n-2)}\right\} \right) \Psi _{\left\{ J_{n-2}\right\}
}\left( \left\{ \theta _f^{(n-2)}\right\} \right) \left( \xi _f\xi _i\eta
_f\eta _i\right) ^{\frac{3-n}2}  \nonumber \\
&&\ \ \!\!\!\!\!\!\!\!\!\!\!\!\!\times \left( \frac{M\omega }{i\hbar }%
\right) ^2\int_0^\infty dS\frac{e^{2\frac i\hbar \gamma S}}{\sin ^2(\omega S)%
}\exp \left[ \frac{iM\omega }{2\hbar }\left( \xi _f^2+\eta _f^2+\xi
_i^2+\eta _i^2\right) \cot (\omega S)\right]   \nonumber \\
&&\ \ \!\!\!\!\!\!\!\!\!\!\!\!\!\times I_{\sigma +\frac{n-3}2}\left( \frac{%
M\omega \xi _f\xi _i}{i\hbar \sin (\omega S)}\right) I_{\sigma +\frac{n-3}%
2}\left( \frac{M\omega \eta _f\eta _i}{i\hbar \sin (\omega S)}\right)  
\nonumber \\
\!\!\!\!\!\!\!\!\!\!\!\!\!\!\!\!\!\!\!\!\!\!\!\!\!\!\!\!\!\!\!\!\!\!\!\!\!\!%
\!\!\!\!\!\!\!\!\!\!\!\!\!\!\!\!\! &=&\!\!\!\!i\hbar \!\!\!\stackunder{%
N_1,N_2,\left\{ J_{n-2}\right\} }{\sum }\!\!\!\!\!\!\!\!\!\!\frac{\Psi
_{N_1,N_2,\left\{ J_{n-2}\right\} }^{*}\!\!\left( \!\xi _i,\eta _i,\!\left\{
\theta _i^{(n-2)}\right\} \!\right) \!\Psi _{N_1,N_2,\left\{ J_{n-2}\right\}
}\!\!\left( \!\xi _f,\eta _f,\!\!\left\{ \theta _f^{(n-2)}\right\} \!\right) 
}{E-E_{N_1,N_2,\left\{ J_{n-2}\right\} }},  \nonumber  \label{a.23} \\
&&  \label{a.25}
\end{eqnarray}
where the properly normalized wave functions and the energy spectrum are
given by

\begin{eqnarray}
\!\!\!\!\!\!\!\!\!\!\!\!\!\Psi _{N_1,N_2,\left\{ J_{n-2}\right\} }\left( \xi
,\eta ,\left\{ \theta ^{(n-2)}\right\} \right) \!\!\!\! &=&\!\!\frac 1{\sqrt{%
N}}\left( \!\frac{N_1!N_2!}{\Gamma \left( N_1+\sigma +\frac{n-1}2\right)
\Gamma \left( N_2+\sigma +\frac{n-1}2\right) }\!\right) ^{\frac 12} 
\nonumber \\
&&\ \!\!\!\!\times \left( -\frac{2ME_N}{\hbar ^2}\right) ^{\frac{n+2\sigma }%
4}(\xi \eta )^\sigma \exp \left[ -\sqrt{-\frac{2ME_N}{\hbar ^2}}\frac{\xi
^2+\eta ^2}2\right]   \nonumber \\
&&\ \!\!\!\!\!\!\!\!\times L_{N_1}^{\sigma +\frac{n-3}2}\left( \!\!\sqrt{-%
\frac{2ME_N}{\hbar ^2}}\xi ^2\!\!\right) L_{N_2}^{\sigma +\frac{n-3}2}\left(
\!\!\sqrt{-\frac{2ME_N}{\hbar ^2}}\eta ^2\right) \!\!  \nonumber \\
&&\ \times \Psi _{\left\{ J_{n-2}\right\} }\left( \left\{ \theta
^{(n-2)}\right\} \right) ,  \label{a.26}
\end{eqnarray}
\begin{equation}
E_N=E_{N_1,N_2,\left\{ J_{n-2}\right\} }=-\frac{M\gamma ^2}{2\hbar ^2N^2},
\label{a.27}
\end{equation}
with $N=N_1+N_2+\sigma +\frac{n-1}2.$ The energy levels and wave functions
have been obtained by means of the Hille-Hardy formula (\cite{ISGIMR} ,
p.1038):

\begin{equation}
\!\!\!\!\stackunder{n=0}{\stackrel{\infty }{\sum }}\frac{n!}{\Gamma
(n+\alpha +1)}L_n^\alpha (x)L_n^\alpha (y)z^n\!\!=\!\!\frac{(xyz)^{-\frac
\alpha 2}}{1-z}\exp \left( -z\frac{x+y}{1-z}\right) \!I_\alpha \!\left( \!%
\frac{2\sqrt{xyz}}{1-z}\right) ;\left| z\right| \!<1,  \label{a.28}
\end{equation}
where $L_n^\alpha (x)$ are the Laguerre polynomials (\cite{ISGIMR} , p.1037).

\section{Spherical coordinates in E$_n$ and path integral}

We shall now discuss the path integral for this class of super-integrable
systems in spherical coordinates. In $n$-dimensional Euclidean space E$_n,$
the cartesian coordinates ( $x^{(1)},x^{(2)},...,x^{(n)}$), are related to
the spherical coordinates ($r,\theta ^{(1)},\theta ^{(2)},...,\theta
^{(n-1)} $), by means of the following transformation: 
\begin{eqnarray}
x^{(1)} &=&r\cos \theta ^{(1)}  \nonumber \\
x^{(2)} &=&r\sin \theta ^{(1)}\cos \theta ^{(2)}  \nonumber \\
&&.  \nonumber \\
&&.  \nonumber \\
&&.  \nonumber \\
x^{(n-1)} &=&r\sin \theta ^{(1)}...\sin \theta ^{(n-2)}\cos \theta ^{(n-1)} 
\nonumber \\
x^{(n)} &=&r\sin \theta ^{(1)}...\sin \theta ^{(n-2)}\sin \theta ^{(n-1)},
\label{a.29}
\end{eqnarray}
with $0<r<\infty ,0\leq \theta ^{(k)}<\pi ,k=1,2,...,n-2,0\leq \theta
^{(n-1)}<2\pi $ and $r^2=\stackunder{k=1}{\stackrel{n}{\sum }}\left(
x^{(k)}\right) ^2.$ In these spherical coordinates the classical Lagrangian
has the form:

\begin{eqnarray}
\!\!\!\!\!\!\!\!\mathcal{L}_{cl}\!\!\!\! &=&\!\!\!\!\frac M2\left[ \stackrel{%
.}{r}^2\!\!+r^2\left( \stackrel{.}{(\theta ^{(1)}})^2+\sin ^2\theta ^{(1)}%
\stackrel{.}{(\theta ^{(2)}})^2+...+\sin ^2\theta ^{(1)}...\sin ^2\theta
^{(n-2)}(\stackrel{.}{\theta }^{(n-1)})^2\right) \right]   \nonumber \\
&&\ \!\!\!\!\!\!\!\!\!\!\!+\!\frac \gamma r-\frac{\hbar ^2}{2Mr^2}\left[ 
\frac{p_1(p_1-1)}{\cos ^2\theta ^{(1)}}+\!\!\frac 1{\sin ^2\theta
^{(1)}}\left[ \frac{p_2(p_2-1)}{\cos ^2\theta ^{(2)}}+\!\!\frac 1{\sin
^2\theta ^{(2)}}\left[ \frac{p_3(p_3-1)}{\cos ^2\theta ^{(3)}}+...\right.
\right. \right.   \nonumber \\
&&\ +\left. \left. \left. \frac{p_{n-1}(p_{n-1}-1)}{\sin ^2\theta
^{(n-2)}\cos ^2\theta ^{(n-1)}}\right] \right] ...\right] ,  \label{a.30}
\end{eqnarray}
and the metric tensor is given by 
\begin{equation}
\left( g_{ab}\right) =diag\left( 1,r^2,g_{\theta ^{(1)}\theta
^{(1)}},...,g_{\theta ^{(n-2)}\theta ^{(n-2)}}\right) ,  \label{a.31}
\end{equation}
where 
\begin{equation}
g_{\theta ^{(i)}\theta ^{(i)}}=\stackunder{k=1}{\stackrel{i}{\prod }}\sin
^2\theta ^{(k)},i=1,2,...,n-2.  \label{a.32}
\end{equation}
Of course 
\begin{equation}
\left( g^{ab}\right) =\left( g_{ab}\right) ^{-1}=diag\left( 1,\frac
1{r^2},g_{\theta ^{(1)}\theta ^{(1)}}^{-1},...,g_{\theta ^{(n-2)}\theta
^{(n-2)}}^{-1}\right) ,  \label{a.33}
\end{equation}
and 
\begin{equation}
\sqrt{g}=\sqrt{\det \left( g_{ab}\right) }=r^{n-1}\stackunder{k=1}{\stackrel{%
n-1}{\prod }}\left( \sin \theta ^{(k)}\right) ^{n-1-k}.  \label{a.34}
\end{equation}
The momentum operators have the form 
\begin{eqnarray}
P_r &=&\frac \hbar i\left( \frac \partial {\partial r}+\frac{\Gamma _r}%
2\right) ;\Gamma _r=\frac{n-1}r,  \nonumber \\
P_{\theta ^{(k)}} &=&\frac \hbar i\left( \frac \partial {\partial \theta
^{(k)}}+\frac{\Gamma _{\theta ^{(k)}}}2\right) ;\Gamma _{\theta
^{(k)}}=(n-1-k)\cot \theta ^{(k)},k=1,2,...,n-1.  \nonumber  \label{a.33} \\
&&  \label{a.35}
\end{eqnarray}
This gives for the Hamiltonian 
\begin{eqnarray}
\!\!\!\!\!\!\!\!\!\!\!\!\!\!\!\!\!\!H\!\!\!\! &=&\!\!\!\!\!\frac
1{2M}\!\left[ P_r^2\!\!+\!\!\frac 1{r^2}\left[ P_{\theta
^{(1)}}^2\!\!+\!\!\frac 1{\sin ^2\theta ^{(1)}}\left[ P_{\theta
^{(2)}}^2\!\!+\!\!\frac 1{\sin ^2\theta ^{(2)}}\left[ P_{\theta
^{(3)}}^2\!\!+\!\!...\!\!+\!\!\frac 1{\sin ^2\theta ^{(n-2)}}P_{\theta
^{(n-1)}}^2\right] \!\!...\!\!\right] \!\right] \!\right]   \nonumber \\
&&\ \!\!\!\!\!\!\!\!-\frac \gamma r+\frac{\hbar ^2}{2Mr^2}\!\left[ \frac{%
p_1(p_1-1)}{\cos ^2\theta ^{(1)}}+\frac 1{\sin ^2\theta ^{(1)}}\!\left[ 
\frac{p_2(p_2-1)}{\cos ^2\theta ^{(2)}}+\frac 1{\sin ^2\theta
^{(2)}}\!\left[ \frac{p_3(p_3-1)}{\cos ^2\theta ^{(3)}}+...\right. \right.
\right.   \nonumber \\
&&\ +\left. \left. \left. \frac 1{\sin ^2\theta ^{(n-2)}}\frac{%
p_{n-1}(p_{n-1}-1)}{\cos ^2\theta ^{(n-1)}}\right] \right] ...\right]
+\triangle V.  \label{a.36}
\end{eqnarray}
Here the potential $\triangle V$ is given by 
\begin{eqnarray}
\triangle V &=&\frac{\hbar ^2}{8Mr^2}\left[ (n-1)(n-3)+\frac 1{\sin ^2\theta
^{(1)}}\left[ (n-2)^2\cos ^2\theta ^{(1)}-2(n-2)\right. \right.   \nonumber
\\
&&\ +\frac 1{\sin ^2\theta ^{(2)}}\left[ (n-3)^2\cos ^2\theta
^{(2)}-2(n-3)+\frac 1{\sin ^2\theta ^{(3)}}\left[ (n-4)^2\cos ^2\theta
^{(3)}\right. \right.   \nonumber \\
&&\ \left. \left. \left. \left. -2(n-4)+...+\frac 1{\sin ^2\theta
^{(n-2)}}\left[ \cos ^2\theta ^{(n-2)}-2\right. \right] ...\right] \right]
\right] .  \label{a.37}
\end{eqnarray}

Constructing the path integral in E$_n$, we follow the prescription adopted
in the section II and obtain

\begin{eqnarray}
K(\stackrel{\rightarrow }{r}_f,\stackrel{\rightarrow }{r}_i;T) &=&%
\stackunder{N\rightarrow \infty }{\lim }\int \stackunder{j=1}{\stackrel{N}{%
\prod }}\left( \frac M{2i\pi \hbar \varepsilon }\right) ^{\frac n2}%
\stackunder{j=1}{\stackrel{N-1}{\prod }}r_j^{n-1}dr_j\stackrel{n-1}{%
\stackunder{k=1}{\prod }}\left( \sin \theta _j^{(k)}\right) ^{n-1-k}d\theta
_j^{(k)}  \nonumber \\
&&\ \times \exp \left\{ \frac i\hbar \stackunder{j=1}{\stackrel{N}{\sum }}%
S^{sc}(j,j-1)\right\} ,  \label{a.38}
\end{eqnarray}
where 
\begin{eqnarray}
S^{sc}(j,j-1) &=&\frac M{2\varepsilon }\left[ \triangle ^2r_j+\widehat{r}%
_j^2\triangle ^2\theta _j^{(1)}+\widehat{r}_j^2\widehat{\sin ^2\theta
_j^{(1)}}\triangle ^2\theta _j^{(2)}\right.  \nonumber \\
&&\ \left. +\widehat{r}_j^2\widehat{\sin ^2\theta _j^{(1)}}...\widehat{\sin
^2\theta _j^{(n-2)}}\triangle ^2\theta _j^{(n-1)}\right] +\frac{\varepsilon
\gamma }{r_j}  \nonumber \\
&&\ -\frac{\varepsilon \hbar ^2}{2Mr_j^2}\left[ \frac 14(n-1)(n-3)+\frac{%
p_1(p_1-1)}{\cos ^2\theta _j^{(1)}}\right.  \nonumber \\
&&\ +\frac 1{\sin ^2\theta _j^{(1)}}\left[ \frac{p_2(p_2-1)}{\cos ^2\theta
_j^{(2)}}+\frac{(n-2)^2\cos ^2\theta _j^{(1)}-2(n-2)}4\right.  \nonumber \\
&&\ +\frac 1{\sin ^2\theta _j^{(2)}}\left[ \frac{p_3(p_3-1)}{\cos ^2\theta
_j^{(3)}}+\frac{(n-3)^2\cos ^2\theta _j^{(2)}-2(n-3)}4\right.  \nonumber \\
&&\ \left. \left. \left. +...\!+\!\frac 1{\sin ^2\theta _j^{(n-2)}}\left[ \!%
\frac{p_{n-1}(p_{n-1}-1)}{\cos ^2\theta _j^{(n-1)}}\!+\!\frac{\cos ^2\theta
_j^{(n-2)}-2}4\!\right] \!...\!\right] \right] \right] .  \nonumber
\label{a.37} \\
&&  \label{a.39}
\end{eqnarray}

Path integration in spherical coordinates is quite similar to that carried
out in parabolic rotational coordinates. By performing successive time
transformations $dt=r^2\stackrel{n-1}{\stackunder{k=i}{\prod }}\sin ^2\theta
^{(n-1-k)}ds;(i=1,2,...,n-2),$ and their inverses, we obtain in the
separation steps for the angular variables:

\begin{equation}
K(\stackrel{\rightarrow }{r}_f,\stackrel{\rightarrow }{r}_i;T)=\stackrel{%
\infty }{\stackunder{\left\{ J_{n-1}\right\} =0}{\sum }}\!\Phi _{\left\{
J_{n-1}\right\} }^{*}\left( \!\left\{ \theta _i^{(n-1)}\right\} \!\right)
\Phi _{\left\{ J_{n-1}\right\} }\left( \!\left\{ \theta _f^{(n-1)}\right\}
\!\right) \!K_{\left\{ J_{n-1}\right\} }(r_f,r_i;T),  \label{a.40}
\end{equation}
where $\left\{ J_{n-1}\right\} $ and $\left\{ \theta ^{(n-1)}\right\} $
denote the sets $\left\{ J_1,J_2,...,J_{n-1}\right\} $ and \\$\left\{ \theta
^{(1)},\theta ^{(2)},...,\theta ^{(n-1)}\right\} ,$ respectively. The
angular part of the wave functions is given by 
\begin{equation}
\Phi _{\left\{ J_{n-1}\right\} }\left( \left\{ \theta ^{(n-1)}\right\}
\right) =\stackrel{n-1}{\stackunder{k=1}{\prod }}\Phi _{J_k}(\theta ^{(k)}),
\label{a.41}
\end{equation}
where 
\begin{eqnarray}
\Phi _{J_k}(\theta ^{(k)}) &=&\left[ \frac{2(m_{k+1}+p_k+2J_k+\frac
12)J_k!\Gamma (m_{k+1}+p_k+J_k+\frac 12)}{\Gamma (p_k+J_k+\frac 12)\Gamma
(m_{k+1}+J_k+1)}\right] ^{\frac 12}  \nonumber \\
&&\ \times \left( \sin \theta ^{(k)}\right) ^{m_{k+1}+1-\frac{n-k}2}\left(
\cos \theta ^{(k)}\right) ^{p_k}P_{J_k}^{(m_{k+1},p_k-\frac 12)}(\cos
2\theta ^{(k)}),  \nonumber  \label{a.40} \\
&&  \label{a.42}
\end{eqnarray}
with 
\begin{equation}
m_n=-\frac 12;\quad m_k=\stackrel{n-1}{\stackunder{i=k}{\sum }}p_i+2%
\stackrel{n-1}{\stackunder{i=k}{\sum }}J_i+\frac{n-k-1}2;\quad k=1,2,...,n-1.
\label{a.43}
\end{equation}
The radial path integral $K_{\left\{ J_{n-1}\right\} }(r_f,r_i;T)$ reads as

\begin{eqnarray}
\!\!\!\!\!\!\!\!\!\!\!\!\!\!\!\!\!\!\!K_{\left\{ J_{n-1}\right\}
}(r_f,r_i;T)\!\!\! &=&\!\!\!\frac 1{(r_fr_i)^{\frac{n-1}2}}\int \mathcal{D}%
r(t)\exp \left\{ \frac i\hbar \int_0^T\left[ \!\frac{M\stackrel{.}{r}^2}2-%
\frac{\hbar ^2}{2Mr^2}\left( m_1^2-\frac 14\right) \!\!+\!\!\frac \gamma
r\!\right] dt\right\} .  \nonumber \\
&&  \label{a.44}
\end{eqnarray}

By using the path integral solution for the radial part of the problem of
the Coulomb potential\cite{LCTFH,FS,CG2} , the radial Green function is
evaluated to be

\begin{eqnarray}
\!\!\!\!\!\!\!\!\!\!\!\!\!\!G_{\left\{ J_{n-1}\right\} }(r_f,r_i;E)
&=&\int_0^\infty dT\exp \left( \frac{iE}\hbar T\right) K_{\left\{
J_{n-1}\right\} }(r_f,r_i;T)  \nonumber \\
\!\!\!\!\!\!\!\!\!\!\!\!\!\!\!\!\!\!\!\!\!\!\!\!\!\!\!\!\!\!\!\!\!\!\!\!\!\!%
\!\!\!\!\!\!\!\!\!\!\!\!\!\!\!\!\!\!\!\! &=&\!\!\!\!\frac{M\omega }{i\hbar
(r_fr_i)^{\frac{n-2}2}}\!\!\!\int_0^\infty \!\!\!dS\frac{e^{\frac{4i\gamma }%
\hbar S}}{\sin (\omega S)}\exp \left\{ \!\frac{iM\omega }{2\hbar }%
(r_f+r_i)\cot (\omega S)\right\}  \nonumber  \label{a.43} \\
&&\ \times I_{2m_1}\left( \frac{M\omega \sqrt{r_fr_i}}{i\hbar \sin (\omega S)%
}\right)  \label{a.45} \\
\!\!\!\!\!\!\!\!\!\!\!\!\!\! &=&\!\!\frac 1{i\omega (r_fr_i)^{\frac{n-1}2}}%
\frac{\Gamma (\kappa +m_1+\frac 12)}{\Gamma (2m_1+1)}W_{-\kappa
,m_1}\!\left( \!\frac{M\omega }\hbar r_f\right) \!M_{-\kappa ,m_1}\!\left( \!%
\frac{M\omega }\hbar r_i\right) ,  \nonumber  \label{a.44} \\
&&  \label{a.46}
\end{eqnarray}
where $\kappa =-2\gamma /\hbar \omega $, $\omega =2(-2E/M)^{1/2}$ and $%
r_f>r_{i\text{ }}.$ The $M_{-\kappa ,m_1}\left( x\right) $ and $W_{-\kappa
,m_1}\left( x\right) $ are the Whittaker functions (\cite{ISGIMR} , p. 1059).

Expanding the radial Green function by means of the Hille-Hardy formula (\ref
{a.28}), we obtain after performing the integration over $S$ in Eq.(\ref
{a.45}) the energy spectrum and the bound state wave functions, respectively

\begin{equation}
E_N=E_{N_r,\left\{ J_{n-1}\right\} }=-\frac{M\gamma ^2}{2\hbar ^2N^2};\quad
N=N_r+m_1+\frac 12,  \label{a.47}
\end{equation}
\begin{eqnarray}
\!\!\!\!\!\Psi _{N_r,\left\{ J_{n-1}\right\} }(r,\left\{ \theta
^{(n-1)}\right\} )\!\!\!\! &=&\!\!\!\!\left[ \!\frac{aN_r!}{4\Gamma
(N_r+2m_1+1)}\!\right] ^{\frac 12}\left( \!\frac 2{a(N_r+m_1+\frac
12)}\!\right) ^{m_1+\frac 32}\!r^{m_1-\frac{n-2}2}  \nonumber \\
&&\ \exp \left( -\frac r{a(N_r+m_1+\frac 12)}\right) L_{N_r}^{2m_1}\left( 
\frac{2r}{a(N_r+m_1+\frac 12)}\right)  \nonumber  \label{a.46} \\
&&\times \Phi _{\left\{ J_{n-1}\right\} }(\left\{ \theta ^{(n-1)}\right\} ),
\label{a.48}
\end{eqnarray}
with $a=\hbar ^2/M\gamma ,$ and $N_r$ is the radial quantum number.

When we put $\beta _i=0$ in the expression of $%
V(x^{(1)},x^{(2)},...,x^{(n)}) $, we obtain the Coulomb potential in $n$%
-dimensional Euclidean space. In this case, our results are reduced to those
obtained by the path integral approach \cite{LCTFH} , or through the
resolution of the Schr\"odinger equation \cite{LCH} .

\section{Conclusion}

In the present work we have shown that the Green function associated to a
family of super-integrable systems containing the hydrogen atom as a special
case can be calculated by path integral approach in $n$-dimensional
Euclidean space. The explicit path integration has been done in two
coordinates systems, namely in parabolic rotational and in spherical
coordinates. The$\left\{ \theta ^{(k)}\right\} $-dependent path integration
was manageable by time transformations yielding path integrals of
P\"oschl-Teller potential forms. It is interesting to note that the
construction of the Green function as a spectral expansion gives
simultaneously the normalized wave functions and the energy spectrum.

\end{document}